\documentclass[useAMS,usenatbib]{mn2e}
\input psfig.sty

\usepackage[english]{babel}
\usepackage{graphicx}
\usepackage{hyperref}
\usepackage{amssymb}
\usepackage{amsmath}
\usepackage{amsfonts}

\newcommand{\den}{\textsf d}
\newcommand{\dg}{\den_{\rm G}}
\newcommand{\dng}{\den_{\rm NG}}

\title[Optimizing the Fisher Information]
{Optimizing the Recovery of Fisher Information in the Dark Matter
Power Spectrum}

\author[J. Harnois-D\'{e}raps et al.]
{Joachim Harnois-D\'{e}raps\thanks{E-mail: jharno@cita.utoronto.ca}$^{1,2}$, Hao-Ran Yu$^{3,1}$, Tong-Jie
Zhang$^{3,4}$ and Ue-Li
Pen$^{1}$\\
$^1$Canadian Institute for Theoretical Astrophysics, University of
Toronto, M5S 3H8, Ontario, Canada\\
$^2$Department of Physics, University of Toronto, M5S 1A7, Ontario,Canada\\
$^3$Department of Astronomy, Beijing Normal University, Beijing, 100875, P. R. China; tjzhang@bnu.edu.cn\\
$^4$Center for High Energy Physics, Peking University, Beijing, 100871, P.R. China\\
}

\begin{document}

\date{}

\maketitle

\label{firstpage}

\begin{abstract}
We combine two Gaussianization techniques -- Wavelet Non-Linear Wiener Filter (WNLWF) and density reconstruction -- to quantify the recovery of Fisher information that is lost in the gravitational collapse. We compute a displacement fields, in analogy with the Zel'dovich approximation, and apply a Wavelet Non-Linear Wiener Filter that decomposes the reconstructed density fields into a Gaussian and a non-Gaussian component. From a series of 200 realizations of $N$-body simulations, we compute the recovery performance for density fields obtained with both dark matter particles and haloes. We find that the height of the Fisher information trans-linear plateau is increased by more than an order of magnitude at $k>1.0 h\mbox{Mpc}^{-1}$  for particles, whereas either technique alone offers an individual recovery boost of only a factor of three to five. We conclude that these two techniques work in a symbiosis, as their combined performance is stronger than the sum of their individual contribution. When applied to the halo catalogues, we find that the reconstruction has only a weak effect on the recovery of Fisher Information, while the non-linear wavelet filter boosts the information by about a factor of five. We also observe that non-Gaussian Poisson noise saturates the Fisher information, and that shot noise subtracted measurements exhibit a milder information recovery.
\end{abstract}


\section{Introduction}\label{s.intro}

Understanding the nature of dark energy has been identified internationally as one of the main goal of modern cosmology \citep{2006astro.ph..9591A}, and many dedicated experiments attempt to constrain its equation of state: LSST\footnote{http://www.lsst.org/lsst/} \citep{2009arXiv0912.0201L}, EUCLID\footnote{http://www.congrex.nl/09c08/} \citep{2010ASPC..430..266B}, JDEM\footnote{http://science.nasa.gov/missions/jdem/} \citep{2010arXiv1008.4936G}, CHIME\footnote{http://www.physics.ubc.ca/chime/} \citep{2006astro.ph..6104P}, SKA\footnote{http://www.skatelescope.org/} \citep{2007HiA....14..539S,2009IEEEP..97.1482D}, BOSS\footnote{http://cosmology.lbl.gov/BOSS/} \citep{2009astro2010S.314S} and Pan-STARRS\footnote{http://pan-starrs.ifa.hawaii.edu/public/}. One of the favored technique involves a detection of the Baryonic Acoustic Oscillations (BAO) signal \citep{Seo:2003pu,Seo:2005ys, Eisenstein:2006nj, Seo:2007ns}, which has successfully constrained the dark energy parameter in current galaxy surveys \citep{ Eisenstein:2005su, 2006PhRvD..74l3507T, 2007MNRAS.381.1053P,  2011arXiv1105.2862B}. The analyses are based on a detection of the BAO wiggles in the matter power spectrum, which act as a standard ruler and allow one to map the cosmic expansion.

With the new and upcoming generation of dark energy experiments, the precision at which we will be able to measure the cosmological parameters is expected to drop at the sub-percent level, therefore it is essential to understand and suppress every sources of systematic uncertainty. In a BAO analysis, one of the main challenge is to extract an optimal and unbiased observed power spectrum, along with its uncertainty; the latter propagates directly on the dark energy parameters with Fisher matrices \citep{Fisher1935,1997ApJ...480...22T}. This task is difficult for a number of reasons.

For instance, the scales that are relevant for the analyses sit at the transition between the linear and the non-linear regime, at least for the redshift at which current galaxy surveys are sensitive, hence the underlying uncertainty on the matter power spectrum is affected by the non-linear dynamics. These effectively couples the phases of different Fourier modes \citep{2003ApJ...598..818Z} and the Gaussian description of the density fields has been observed to fail \citep{1999MNRAS.308.1179M, 2005MNRAS.360L..82R,2006MNRAS.371.1205R,2006MNRAS.370L..66N,2007MNRAS.375L..51N}. For an estimate of the BAO dilation scale to be robust, one must therefore include in the analysis the full non-linear covariance of the power spectrum. Although results from \cite{2011ApJ...726....7T} seem to suggest that non-Gaussianities had no real effect on the final results, it was recently shown that this was only true if the original power spectrum was measured in an unbiased and optimal way, which is rarely the case \citep{2011arXiv1106.5548N}. Otherwise, the discrepancy on the constraining power is at the percent level. One of the way to reduce the impact of the non-linear dynamics is to transform the observed field into something that is more linear. Over the last few years, many ``Gaussianization'' techniques have been developed, which all attempt to undo the phase coupling between Fourier modes. The number of degrees of freedom -- i.e. uncoupled phases -- can be simply quantified by the Fisher information, and recovering parts of this erased information can lead to improvements by factors of a few on cosmological parameters. 

For example, a density reconstruction algorithm \citep{2007ApJ...664..675E,2009PhRvD..80l3501N,2009PhRvD..79f3523P}, based on the Zel'dovich approximation, has been shown to reduce by a factor of two the constraints on the BAO dilation scale \citep{Eisenstein:2006nk,2011arXiv1106.5548N}. This technique was applied on the SDSS data recently 
\citep{2012arXiv1202.0090P} to improve the BAO detection, with small modifications to the algorithm such as to correct for the survey selection function 
and redshift space distortions. As discussed therein, an important issue is that two main mechanism are reducing our ability to measure the BAO ring accurately:
1) a large coherent $\sim 50$ Mpc infall of the galaxies on to overdensities, which tends to widen the BAO peak, and 
2) local non-linear effects, including non-linear noise, which also erase the smallest BAO wiggles. 
Reconstruction addresses the first of these mechanisms, and its is important to know whether something can be done about the second, 
after reconstruction has been applied.
 
Wavelet Non-linear Wiener Filters (hereafter WNLWF, or just wavelet filter) were used to decompose dark matter density fields \citep{2011ApJ...728...35Z} and weak gravitational lensing $\kappa$-fields \citep{2012MNRAS.421..832Y} into Gaussian and non-Gaussian parts, such as to condense in the latter most of the collapsed structure; the Gaussian part was then shown to contain several times more Fisher information than the original field. Other methods include log-normal transforms \citep{2011ApJ...729L..11S}, Cox-Box \citep{2011arXiv1104.1399J}, running $N$-body simulation backwards \citep{2000ASPC..201..282G}, or direct Gaussianization of the one-point probability function \citep{2011PhRvD..84b3523Y}, just to name a few. This technique seems perfectly suited to address the issue of non-Gaussian noise
described above.

Our focus, in this paper, is to discuss how two of these techniques can be used in conjunction to maximize the recovery of Fisher information. Not all combinations of Gaussianization techniques are winning, however. It was recently shown \citep{2012MNRAS.421..832Y} that WNLWF and log-transforms are not combining in an advantageous way. On one hand, if the log-transform is applied onto a Gaussianized field, the prior on the density field is no longer valid, and the log-transform maps the density into something even less Gaussian. On the other hand, it was shown that the log-transform is less effective than WNLWF alone at recovering Fisher information, at least on small scales. Applying the filter after the log-transform  does not improve the situation, since the Gaussian/non-Gaussian decomposition is less effective. In other words, the Fisher information that the log-transform could not extract is not recovered by WNLWF, and we are better off with the WNLWF alone.

It seems, however, that this unfortunate interaction is not a constant across all combinations. In this paper, we discuss how non-linear Wiener filters, constructed in Wavelet space, can improve the results of a density reconstruction algorithm, which takes the density back in time using linear perturbation theory. Our first result is that these two techniques work well together, in the sense that the final Fisher information recovery is larger  than the two techniques stand alone. We first obtain these results with particle catalogues extracted from $N$-body simulations, and extend our techniques to halo catalogues, which provide a sampling of the underlying matter field that is much closer to actual observations.

The structure of the paper is as follows: in Section \ref{s.theory}, we briefly review the theoretical background of the density reconstruction and WNLWF, and review how we extract the density power spectra, their covariance matrices, and the Fisher information. We discuss our results in Section \ref{s.results} and conclude in Section \ref{s.conclusion}. 

\section{Theoretical Background}
\label{s.theory}

\subsection{Numerical Simulations}
\label{ss.sims}

Our sample of 200 $N$-body simulations are generated with {\small CUBEP3M}, an enhanced version of {\small PMFAST} \citep{2005NewA...10..393M} that solves Poisson equation with sub-grid resolution, thanks to the ${\tt p}^{3}{\tt m}$ calculation. Each run evolves $512^3$ particles on a $1024^{3}$ grid, and is computed on a single IBM node of the Tightly Coupled System on SciNet \citep{Scinet} with $\Omega_{\rm M } = 0.279$, $\Omega_{\Lambda} = 0.721$, $\sigma_8 = 0.815$, $n_s = 0.96$ and $h = 0.701$. We assumed a flat universe, and started the simulations at $z_i = 50$. 
Each simulation has a side of $322.36 h^{-1}\mbox{Mpc}$, and we output the particle catalogue at $z =0.054$.
 We search for haloes with a spherical over-density algorithm \citep{1996MNRAS.281..716C} executed at run time, 
 which sorts the local grid density maxima in descending order of peak height, then loops over the cells surrounding the peak center
 and accumulates the mass until the integrated density drops under the collapse threshold of 178,
 and finally empties the contributing grid cells before continuing with the next candidate, ensuring that each particle contributes to a single halo. 
 Halo candidates must consist of at least one hundred particles, ensuring the haloes are large and collapsed objects. The  center-of-mass of each halo is calculated
 and used as position, as opposed to its peak location, even though both quantities differ by a small amount. 
 We mention here that algorithms of this kind have the unfortunate consequence to create an exclusion region around each halo candidate,
 thus effectively reducing the resolution at which the halo distributions are reliable.
 Each field contains about $88,000$ haloes, for a density of $2.6 \times 10^{-3} h^{3}\mbox{Mpc}^{-3}$. For comparison, this is about eight  times
larger than the density of the BOSS density \citep{2009astro2010S.314S}.

\subsection{Density reconstruction algorithm}
\label{ss.reco}

We use a density reconstruction algorithm that is based on the linear theory prediction, first found by Zel'dovich \citep{1970A&A.....5...84Z}, that couples the density field $\delta({\bf q},t_0)$  to the displacement field ${\bf s}({\bf q})$ via
\begin{eqnarray}
\delta({\bf q},t_0) = - \nabla \cdot {\bf s}({\bf q})
\end{eqnarray}
In the above expression, ${\bf q}$ is the grid, or Lagrangian, coordinate, and the displacement field is obtained in Fourier space as
\begin{eqnarray}
{\bf s}({\bf k}) = -\frac{i{\bf k}}{k^{2}} \delta{(\bf k},t_0)  F(k)
\end{eqnarray}
where $F(k) = \mbox{exp}[-(kR)^{2}/2]$ is a smoothing function suppressing features smaller than $R = 10 h^{-1}\mbox{Mpc}$.
Particles at Eulerian coordinate ${\bf x}$ are displaced from their grid positions following
\begin{eqnarray}
{\bf x}(t_0) = a(t)[{\bf q} + D(t_0)  {\bf s}({\bf q})]
\end{eqnarray}
with $D(t)$ the linear growth factor. These calculations are commonly used for the generation of initial conditions in $N$-body simulations, and are accurate as long as the smallest scales probed are still on linear regime at the starting redshift. In the case where the particles -- or haloes -- to be displaced are not at $t_0$, one must subtract from the result the displacement field from the grid location (see \cite{2009PhRvD..80l3501N} for a detailed explanation of this technique).

\subsection{Wavelet non-linear Wiener filter (WNLWF)}
\label{ss.wavelet}

In this subsection we briefly review the WNLWF algorithm, and direct the reader to  \cite{2011ApJ...728...35Z, 2012MNRAS.421..832Y} for more details. 

We consider in this paper the Daubechies-4 \citep{Daubechies1992} discrete wavelet transform (DWT), which contains certain families of {\it scaling} functions $\phi$ and {\it difference} functions (or {\it wavelet}
functions) $\psi$. The density fields are expanded into combinations of these orthogonal bases, and weighted by  scaling function coefficients (SFCs) and  wavelet function coefficients (WFCs). In our WNLWF algorithm, we deal with
only WFCs, each of which characterizes the amplitude of the perturbation on a certain wavelength and at a certain locations.

In the three dimensional case, the properties of each perturbation depend on three scale indices $(j_1,j_2,j_3)$ -- controlling the scales of the wavelet Daubechies-4 functions -- and three location indexes $(l_1,l_2,l_3)$ -- controlling their translations. Specifically, on a given dimension, the grid scale corresponding to a specified dilation is  $L/2^j$ ($L=1024$ in our case), and the spatial location is determined by $lL/2^j<x<(l+1)/2^j$. After the wavelet transform, all SFCs and WFCs are stored in a 3-dimensional field, preserving the grid resolution (see \cite{1998World.Sci..2A,nr} for more details).

Our non-linear Wiener filter (NLWF) construction strategy relies on the fact that in wavelet space, the non-Gaussianities are clearly characterized in the probability distribution function (PDF) of the WFCs $\tilde\epsilon_{j_1,j_2,j_3;l_1,l_2,l_3}$. We thus construct our filter by splitting the wavelet transform of the original density, which we label $D$, into a Gaussian ($G$) and a non-Gaussian ($N$) contribution. Since wavelet transforms are linear operations, this  Gaussian/non-Gaussian decomposition happens also in real space when we wavelet transform back the contributions. Namely, we can write in wavelet space and real space respectively,
\begin{eqnarray}\label{dabw}
    D=G+N \mbox{ and } \den=\dg+\dng
\end{eqnarray}
where the original density ($\den$) is expressed as the sum over a Gaussian contribution ($\dg$) and a non-Gaussianized contribution ($\dng$). Our goal is thus to design a filter that concentrates most of the collapsed structure in $\dng$, and thus produces $\dg$ that are closer to linear theory.

The NLWF acts on individual wavelet  {\it modes}, which are defined as combinations (not permutations) of all WFCs having the same three scale indices $(j_1,j_2,j_3)$. For each wavelet mode, the NLWF is determined completely by the PDF $f(x)$  of the corresponding WFCs, which is constructed by looping over the other three indices $(l_1,l_2,l_3)$. We then fit this PDF with the analytical function  presented in \cite{2012MNRAS.421..832Y} (equation (15)):
\begin{eqnarray}\label{eq.fPDF}
    f_{\rm PDF}(x)=\frac{1}{\sqrt{\pi}s^{1-\alpha s^2}}
    \frac{\Gamma(\frac{1}{2}\alpha s^2)}{\Gamma(\frac{1}{2}\alpha s^2-\frac{1}{2})}
    (s^2-x^2)^{-\frac{\alpha s^2}{2}}
\end{eqnarray}
and extract the two parameters $\alpha$ and $s$. These are actually dependent on the second and the fourth central moment of the PDF $f(x)$ $m_2$ and $m_4$, hence we measure the moments first, then extract $\alpha$ and $s$ via:
\begin{eqnarray}\label{eq.get_alpha}
    \alpha=\frac{5m_4-9m_2^2}{2m_2 m_4} \mbox{ and }  s=\sqrt{\left|\frac{2m_2 m_4}{m_4-3m_2^2}\right|},
\end{eqnarray}
We then loop back over all spatial indices $(l_1,l_2,l_3)$ of this wavelet mode and decompose each WFC into two components:
 \begin{eqnarray}\label{eq.alpha}
    w_{\rm G}(x) = -\frac{(\ln f)'(x)}{x} = \left(1+\frac{x^2}{s^2}
    \right)^{-1},
\end{eqnarray}
\begin{eqnarray}\label{eq.beta}
    w_{\rm NG}(x) = 1+\frac{(\ln f)'(x)}{x} = 1-\left(1+\frac{x^2}{s^2}
    \right)^{-1},
\end{eqnarray}
which are functions of $x$ only. Note that the final filter function depends only on $s$, which corresponds to the full width at half maximum of the Gaussian NLWF function $w_{\rm G}$. It characterizes the extent of the departure from a Gaussian PDF: the greater the $s$, the smaller departure is from Gaussian statistics. In comparison, $\alpha$ parameterizes the central deviation of the PDF: $ \sigma_{\rm central}=\alpha^{-\frac{1}{2}}$. The same decomposition is performed on the reconstructed density fields and on those obtained from  the halo catalogues. In this paper, we do not make use of the information contained in the non-Gaussian component and simply discard it,
although it serves as a powerful probe of small scale structures and could help identifying haloes in a (Gaussian) noisy environment. 

\subsection{Information recovery}
\label{ss.fisher}

The calculation of uncertainty about dark energy cosmological parameters is based on a propagation of the uncertainty about the matter power spectrum. In this process, the number of degrees of freedom -- i.e. the Fisher information -- contained in the field is directly related to the constraining power. In this section, we review how the Fisher information about the amplitude of the matter power spectrum is calculated from simulated dark matter particles and haloes.

The power spectrum $P(k)$ of a density contrast $\delta({\bf x})$ is calculated in a standard way:
\begin{eqnarray}
P(k) = \langle P({\bf k}) \rangle = \langle |\delta({\bf k})|^{2}  \rangle
\end{eqnarray}
where the angle brackets refer to an average over our $200$ different realizations and over the solid angle.

The uncertainty about the power spectrum is estimated from a covariance matrix $C$ is defined as
\begin{eqnarray}
    C(k,k')\equiv \frac{1}{N-1}\sum_{i=1}^N
    [ P_i(k)-\langle P(k)\rangle] [P_i(k')-\langle P(k')\rangle],
\end{eqnarray}
where $N$ is the number of realizations and $\langle P(k)\rangle$ is the mean angular power spectrum over all realizations. The cross-correlation coefficient matrix is somehow more convenient to plot since it has higher contrasts, and is defined as:
\begin{eqnarray}\label{rho}
    \rho(k,k')= \frac{C(k,k')}{\sqrt{C(k,k)C(k',k')}}
\end{eqnarray}
The diagonal is normalized to one, and each element represents the degree of correlation between the scales $(k,k')$. In the Gaussian approximation,  the density is completely characterized by the power spectrum. Namely,
\begin{eqnarray}\label{eq:C_gauss}
    C_{\rm G}(k,k') = \frac{2 P^{2}(k)}{N(k)}\delta_{kk'}
\end{eqnarray}
Consequently, $\rho_{\rm G}$ is identical to the identity matrix. As expected from the theory of structure formation, the non-linear collapse of the matter density field tends to couple the Fourier modes, which are otherwise independent, starting from the smallest scales and progressing towards larger scales with time. This coupling is responsible for highly correlated region in $\rho$, and can be understood in terms of higher-order corrections, including the bispectrum, the trispectrum, etc.

The Fisher information measures the number of independent Fourier modes in a density field up to a resolution scale $k_{\rm max}$. We see from equation(\ref{eq:C_gauss}) that dividing the covariance by the (square of the) power spectrum is proportional to the number of independent measurements $N(k)$. Therefore, the normalized covariance is defined as:
\begin{eqnarray}\label{Cnorm}
    C_{\rm norm}(k,k')=\frac{C(k,k')}{P(k)P(k')},
\end{eqnarray}
Then, the number of degrees of freedom up to a scale $k_{\rm max}$ is otained by inverting the corresponding sub-sample of the normalized matrix, then summing over all the elements:
 \begin{eqnarray}\label{cumuinfo}
    I(k_{\rm max})=\sum_{k,k'}^{k_{\rm max}} C_{\rm norm}^{-1}(k,k').
\end{eqnarray}
The inversion of the covariance matrix involved in the calculation of the Fisher information amplifies the noise, hence such measurements typically requires a very strong convergence on the forward matrix. This can otherwise lead to biases of a few percent on derived quantities like the BAO dilation scale \citep{2011arXiv1106.5548N}. Generally, a forward matrix that is closer to diagonal contains more Fisher information; the theoretical maximum corresponds to the Gaussian case, where all modes are independent. Any non-vanishing off-diagonal element reduces the information.




\section{Results}
\label{s.results}

In this section, we describe and quantify our ability at recovering Fisher information with our two Gaussianization techniques, for density fields measured with simulated particles and haloes. We recall that the later is much closer to actual observations, since galaxies trace highly collapsed structures.

\subsection{Density fields}
\label{ss.density}

\begin{figure*}
\includegraphics[width=3.2in]{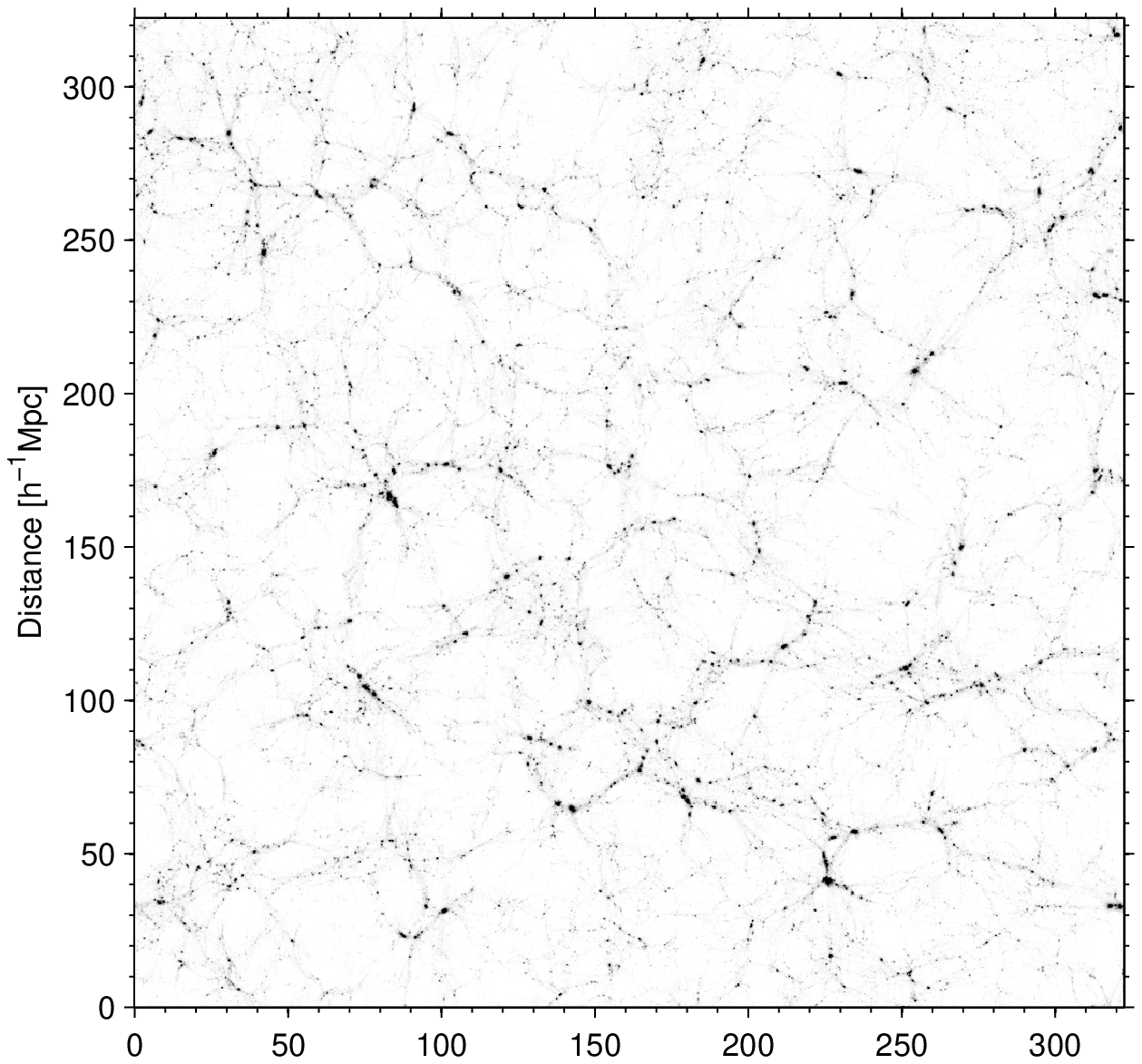}
 \includegraphics[width=3.2in]{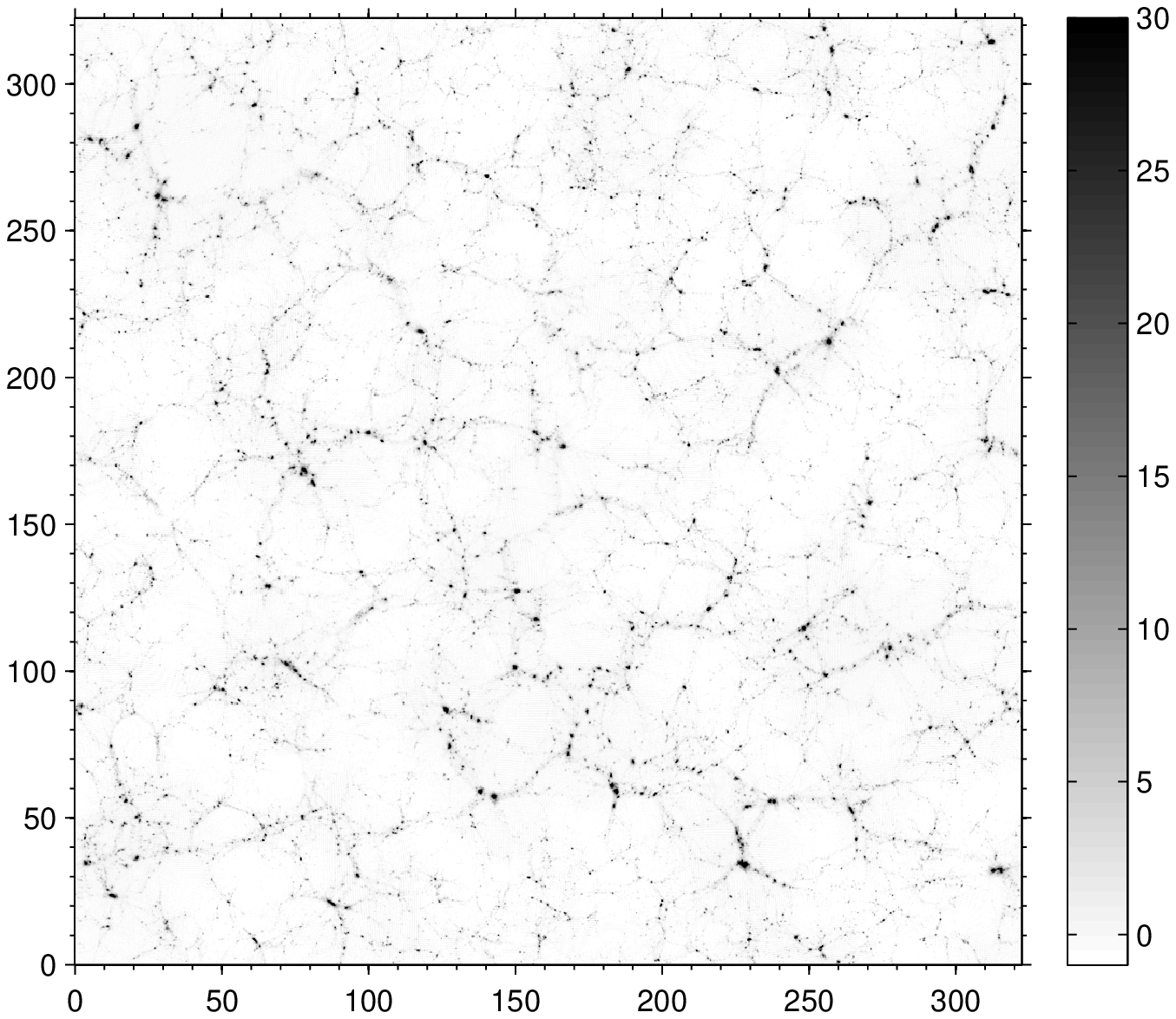}
\includegraphics[width=3.2in]{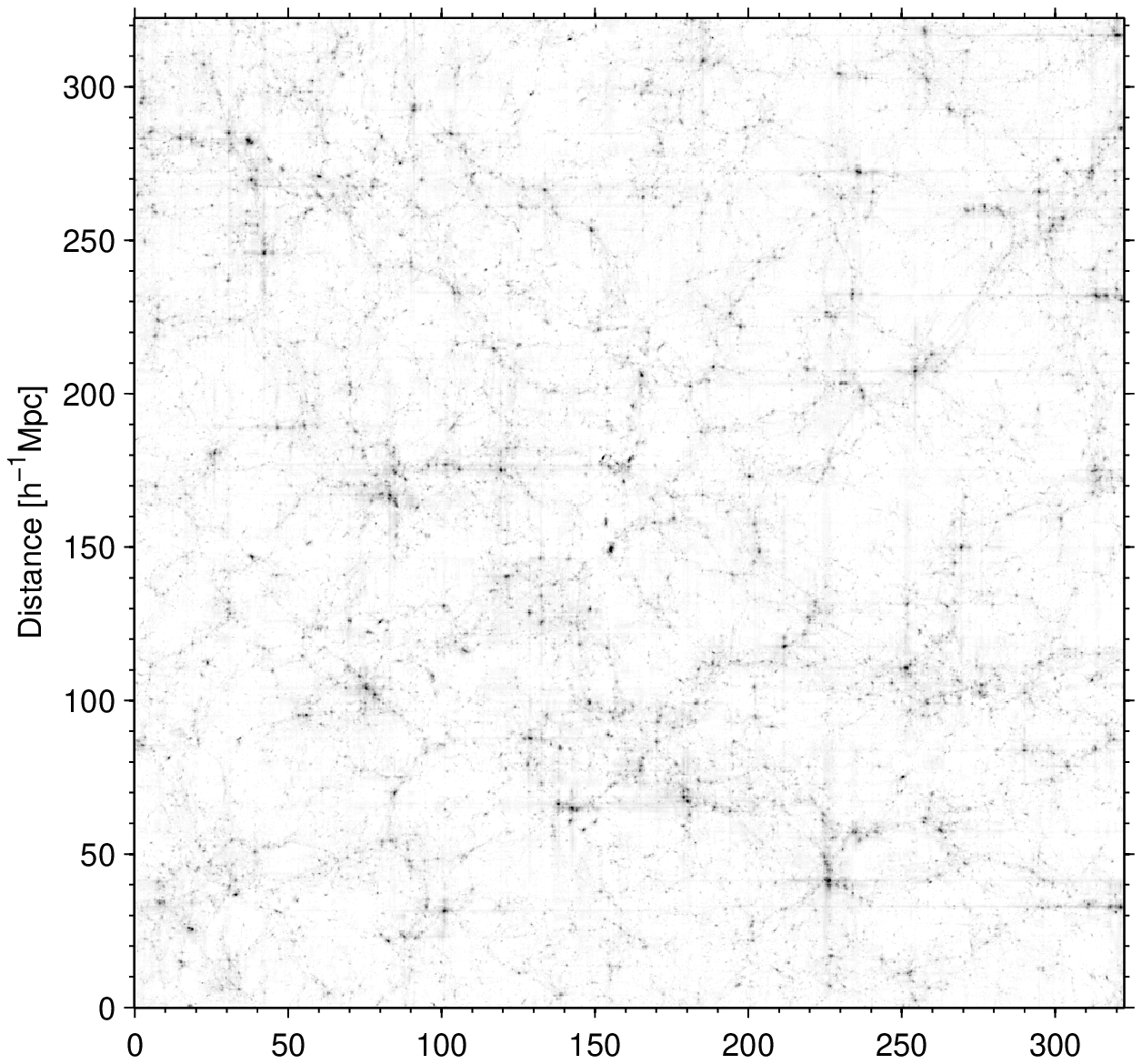}
 \includegraphics[width=3.2in]{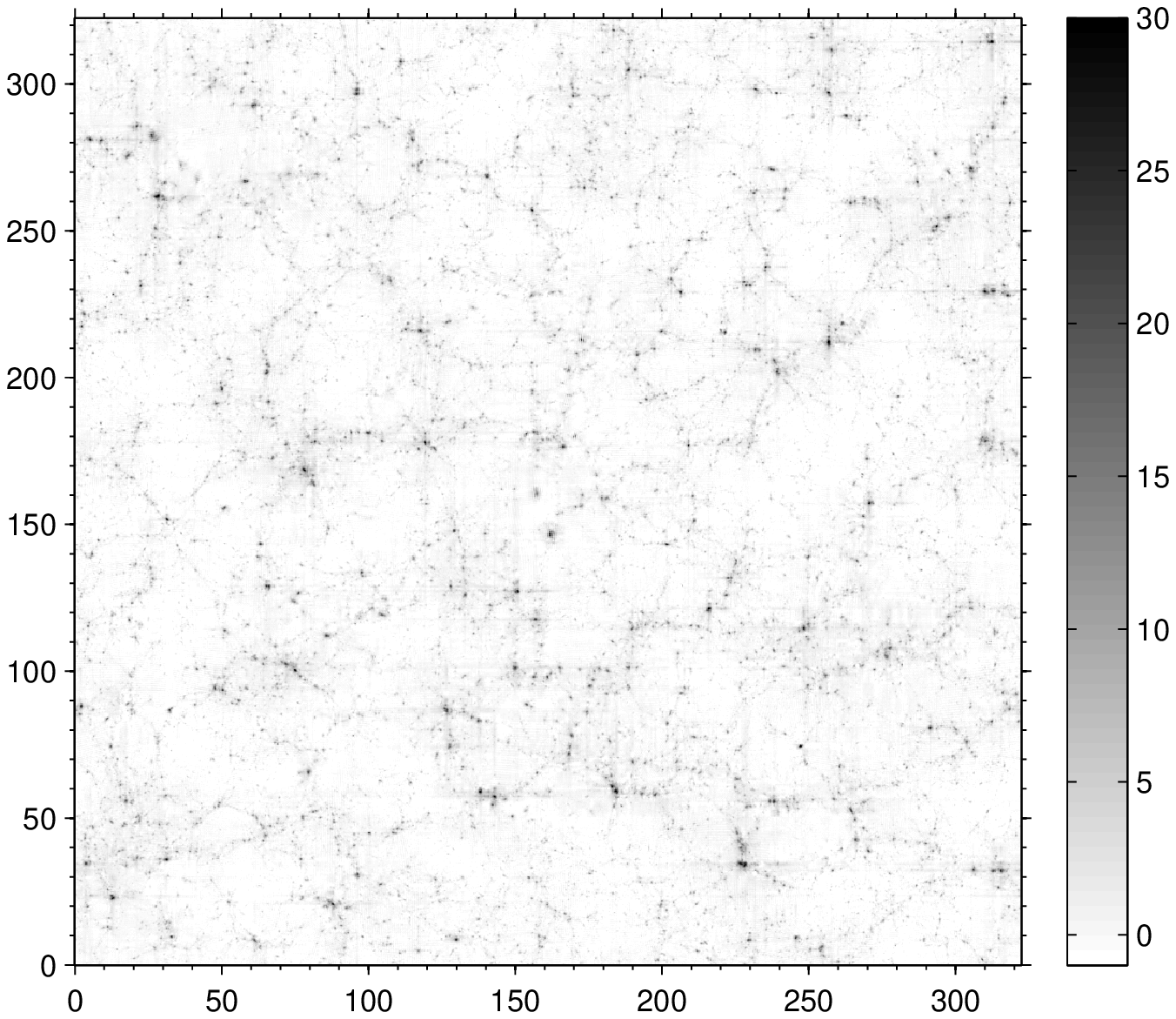}
\includegraphics[width=3.2in]{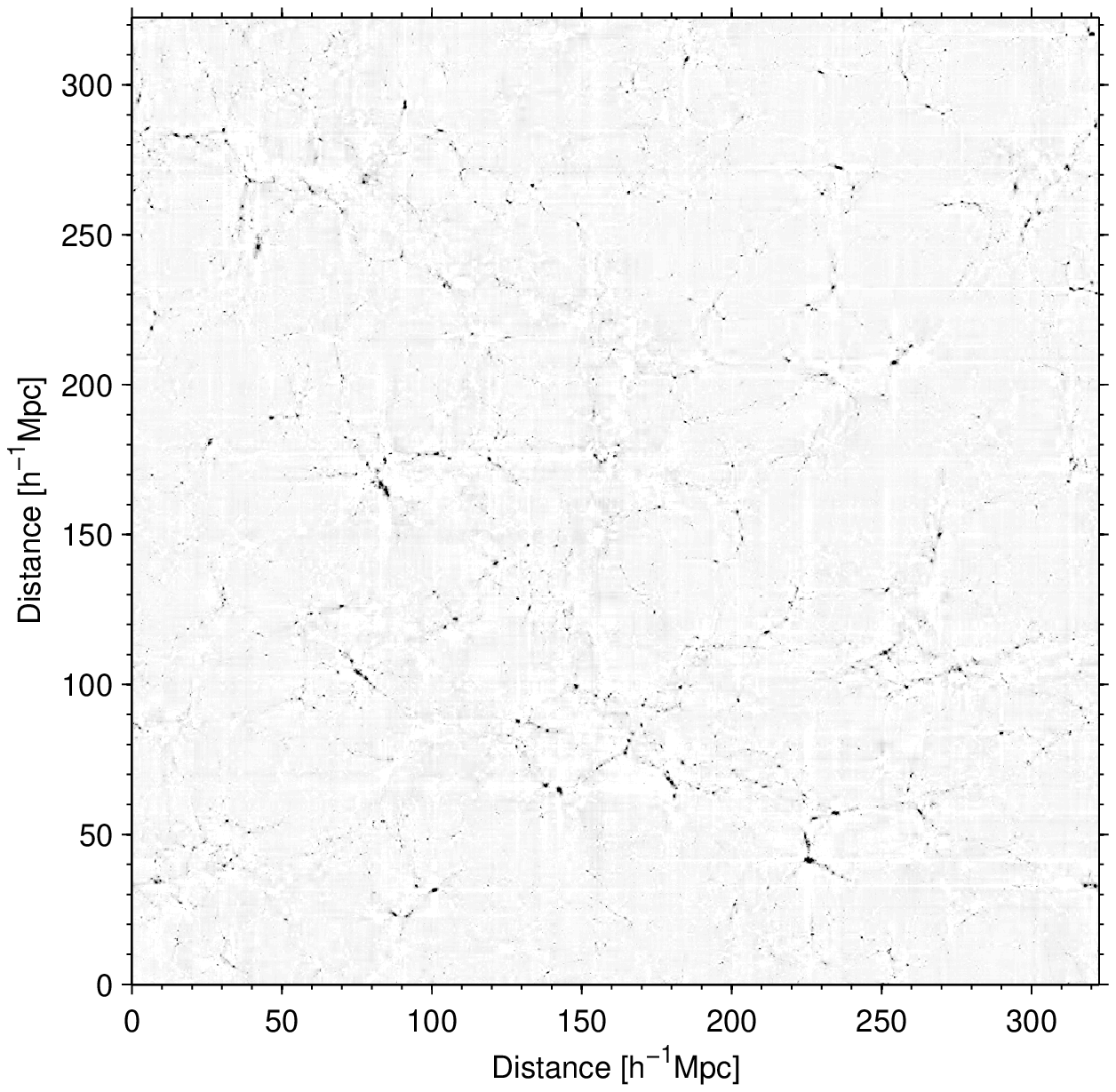}
 \includegraphics[width=3.2in]{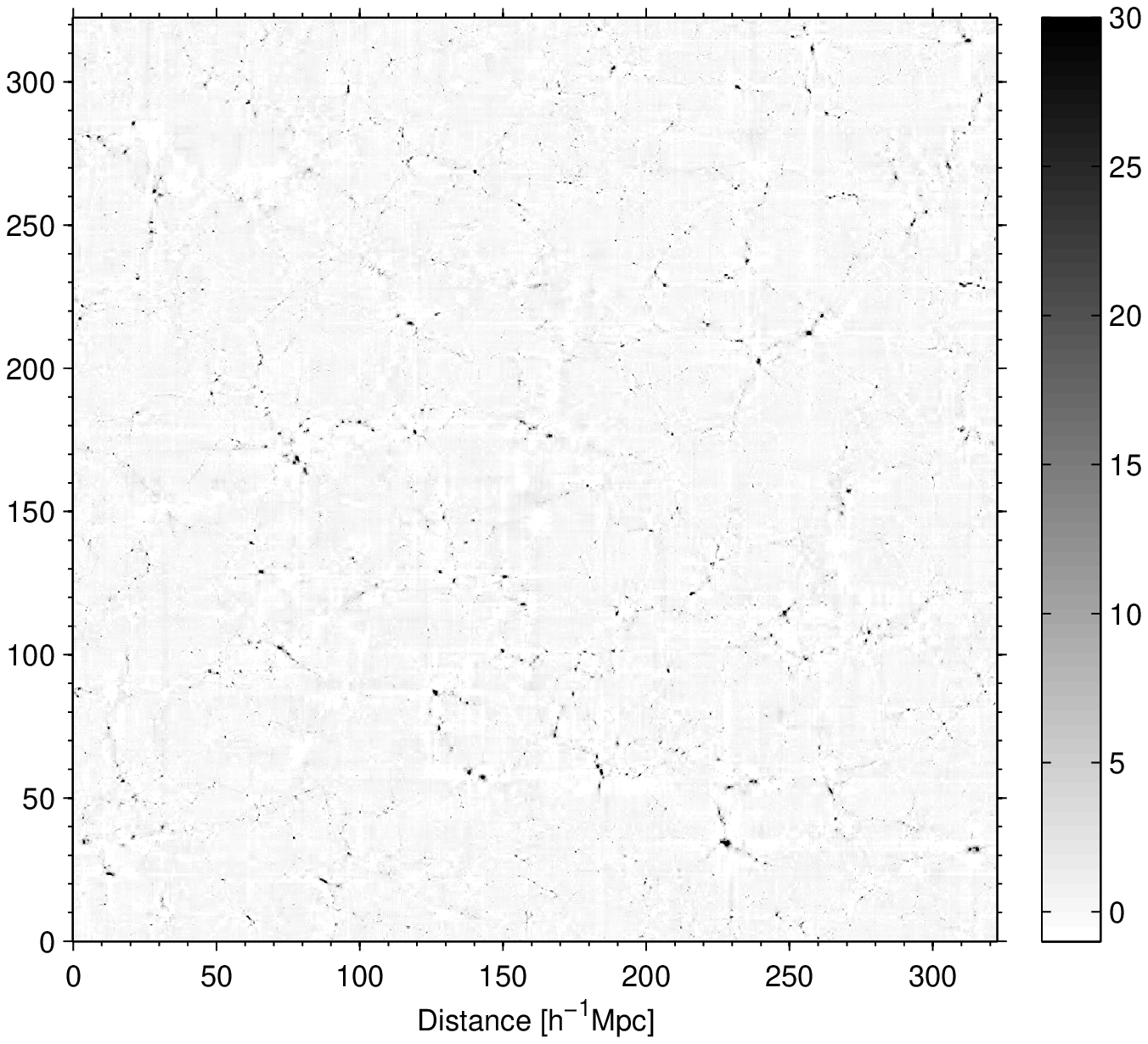}
   \caption{Projections through a thickness of 50 cells of one of the realizations. In each panel, the side is $322.36 h^{-1}\mbox{Mpc}$, and the image contains $1024^2$ pixels. Top left is the original field, top right is the field after linear density reconstruction, middle left is the wavelet filtered field (Gaussian part), middle right is the result of wavelet filtering the reconstructed field. 
   The non-Gaussian part of the wavelet filtered fields are shown in the bottom panels with (right) and without (left) density reconstruction.
   To ease the visual comparison, each panel shows the same overdensity range and saturates for denser regions, i.e. all pixels with $\delta > 30$ are black.}
    \label{f.den}
\end{figure*}

To illustrate the effect of different Gaussianization techniques, we present in Fig. \ref{f.den} the projections through a thickness of 50 cells of a given realization after density reconstruction alone, after WNLWF alone, and with both techniques applied. We observe that the reconstruction reduces the size of each halo, as expected from this algorithm: particles attempt to travel out of the gravitational potential. WNLWF has a slightly different visual effect on the density: it removes most of the smallest structure perturbations, leaving behind the larger ones. As discussed in \cite{1999RSPTA.357.2561P,2011ApJ...728...35Z}, the geometry of the Cartesian wavelet leaves behind a grid patterns, which only affects the smallest scales of the power spectrum and has no impact on the scales we are interested in. The combination of both techniques is presented in the middle right panel, and visually presents the least amount of collapsed structures. We also show the non-Gaussian part of the wavelet filter
with and without reconstruction in the bottom panels. The largest peaks and sharpest structures are indeed filtered out.

\begin{figure*}
\includegraphics[width=6.2in]{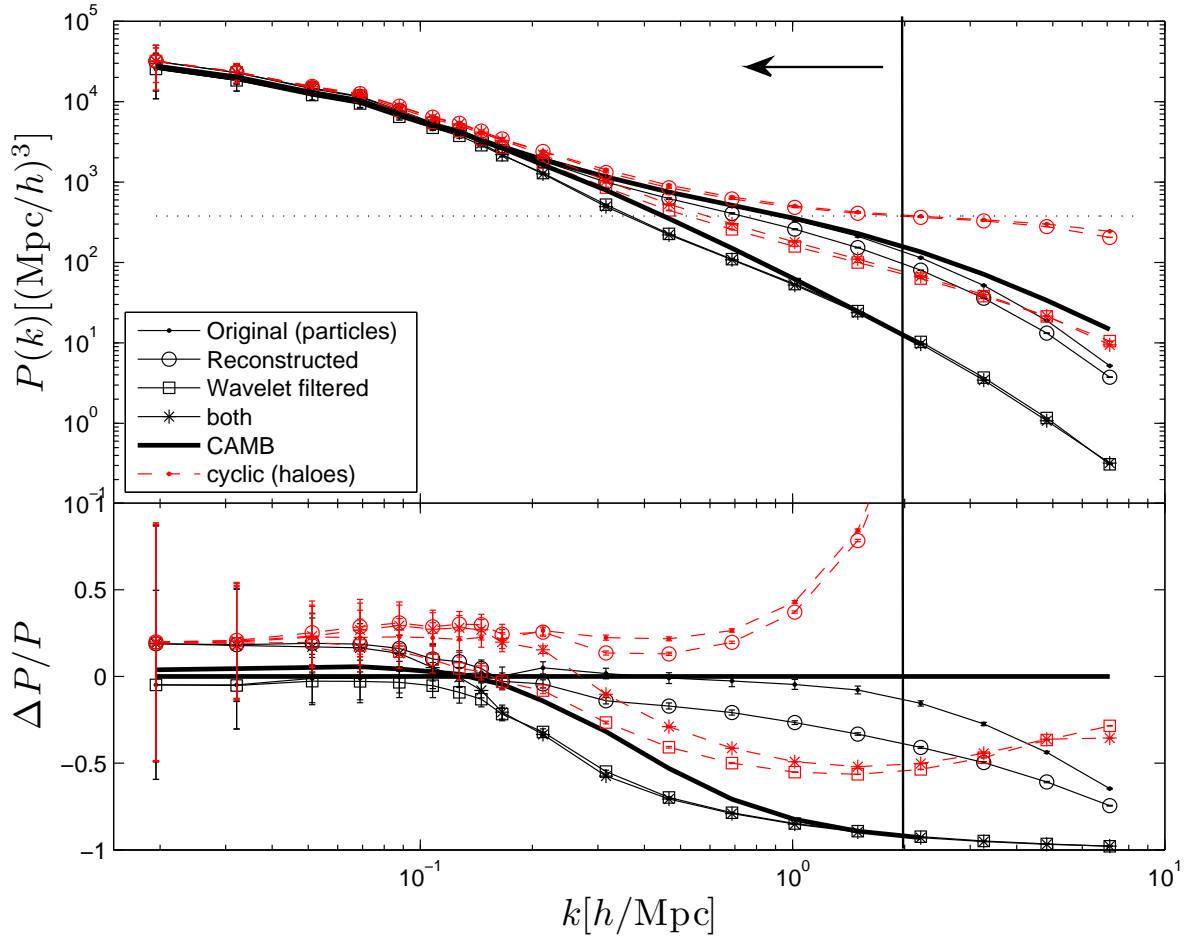}
   \caption{({\it top}:) Power spectra of the original and Gaussianized fields, from simulated particles (symbols + solid  lines) and haloes (symbols + dashed line, in red on the on-line version). The linear and non-linear predictions from {\small CAMB} are shown by the thick solid lines, and the Poisson noise corresponding to the halo population is shown with the thin dotted line. We observe that the wavelet filtered particles densities trace the linear {\small CAMB} predictions, at most within a factor of two. ({\it bottom}:) Fractional error between the curves of the top panel and the non-linear prediction from {\small CAMB}. We observe that the particle power spectrum deviates by more than 10 per cent for $k > 2.0 h\mbox{Mpc}^{-1}$, which sets the resolution limit of our simulations. This scale is represented by the vertical line in both panels. We observe that in linear regime, the wavelet filter preserves the agreement with the predictions, whereas the reconstruction tends to increase the power spectrum by about 20 per cent. This is not a surprise since one effect of the algorithm is to transfer power from small to large scales. In the non-linear regime, however, the power spectrum is highly suppressed by the wavelet filtering process, which factorizes the structures into the non-Gaussian contribution. We measure a linear bias of about 1.2 in all halo measurements. The original and reconstructed halo power spectra are shot noise dominated at scales smaller than $k\sim 1.0 h\mbox{Mpc}^{-1}$, a scale that is strongly suppressed by the wavelet filter.}
    \label{f.power_pp}
\end{figure*}

Fig. \ref{f.power_pp}  shows the power spectrum of the dark matter particles and haloes before and after
the Gaussianization techniques. We first observe that the measurement form  the original particle field agrees at the few percent level with the non-linear predictions obtained from {\small CAMB} \citep{Lewis:1999bs} up to $k\sim 2.0 h\mbox{Mpc}^{-1}$, which sets the resolution scale of our power spectrum measurements. As expected from WNLWF, the Gaussian component of WNLWF preserves the power on linear scales -- up to $k\sim 0.1 h\mbox{Mpc}^{-1}$ -- while signals from trans-linear and non-linear scales are mostly transferred to the non-Gaussian contribution of the WNLWF decomposition, which explains the drop in power. At the same time, we note that the power spectrum after WNLWF actually traces quite well the linear predictions, to within a factor of two, at all scales.

The density reconstruction algorithm also has a significant impact on the shape of the power spectrum, as particles are pumped out of the gravitational potential. As a result, small scale power is directly transferred to larger scales, as seen in the figure. Interestingly, the turning point between these two effects is also locate close to  $k\sim 0.1 h\mbox{Mpc}^{-1}$. This common feature to both Gaussianization technique is explained by the fact that both attempt to pump back information from scales in the trans-linear regime, whose lower-$k$ limit indeed corresponds to this mode. When looking at the halo measurements, we observe that the original and reconstructed power spectra are dominated  by shot noise at scales smaller than $k\sim 1.0 h\mbox{Mpc}^{-1}$. This noise, however, is strongly suppressed by the wavelet filter.
   
\begin{figure}
\includegraphics[width=3.2in]{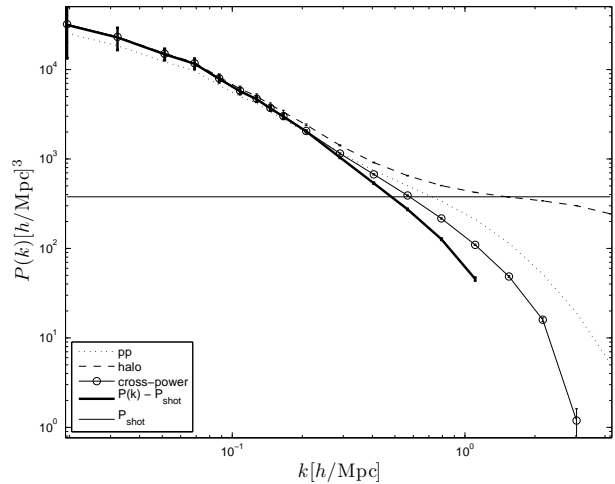}
   \caption{Cross power spectra of a single density field, constructed from a random separation of the haloes onto two distinct fields, whose Fourier transform are combined. The dashed line is the halo power spectrum of the full population, the dotted line is the power spectrum of the particles, the straight line is the Poisson noise estimate, the open symbols represent the cross spectrum of the two randomly selected  populations, and the thick solid line is the shot noise subtracted power. }
    \label{f.cross_power}
\end{figure}

In practice, a common way to deal with Poisson noise is to compute the cross power spectrum between two populations randomly selected out of the original catalogue. The shot noise is eliminated in this operation, and the signals left behind are stronger on large and intermediate scales. Small scales are typically anti-correlated due to the `halo exclusion' effect, which is a result of our halo-finder that collapses all the structure to a single point, and leaves the surrounding region empty. Although precise and robust, this procedure is hard to apply to all the cases under study in this paper, since the density reconstruction algorithm looses efficiency if we under-resolve the small scale structures. Instead, we use another common approach which consists in subtracting from the measured power spectrum an estimate of the shot noise, defined as $P_{\rm shot} = V^{3}/N_{\rm halos}$. To illustrate this, we show in  Fig. \ref{f.cross_power} a comparison between halo power spectrum from the full catalogue, the cross spectrum, and the shot-noise subtracted power. We observe that the two shot noise subtraction techniques agree up to $k = 0.3 h\mbox{Mpc}^{-1}$, beyond which the $P(k) - P_{\rm shot}(k)$ approach looses power in comparison; by $k = 1.0 h\mbox{Mpc}^{-1}$, it is smaller by a factor of 2.4. This difference between the two shot noise subtraction techniques means that our approach is not optimal, and that the results on the smallest scales are not as robust as one would wish.  

\subsection{Covariance matrices}
\label{ss.covmatrix}
\begin{figure*}
    \includegraphics[width=3.2in]{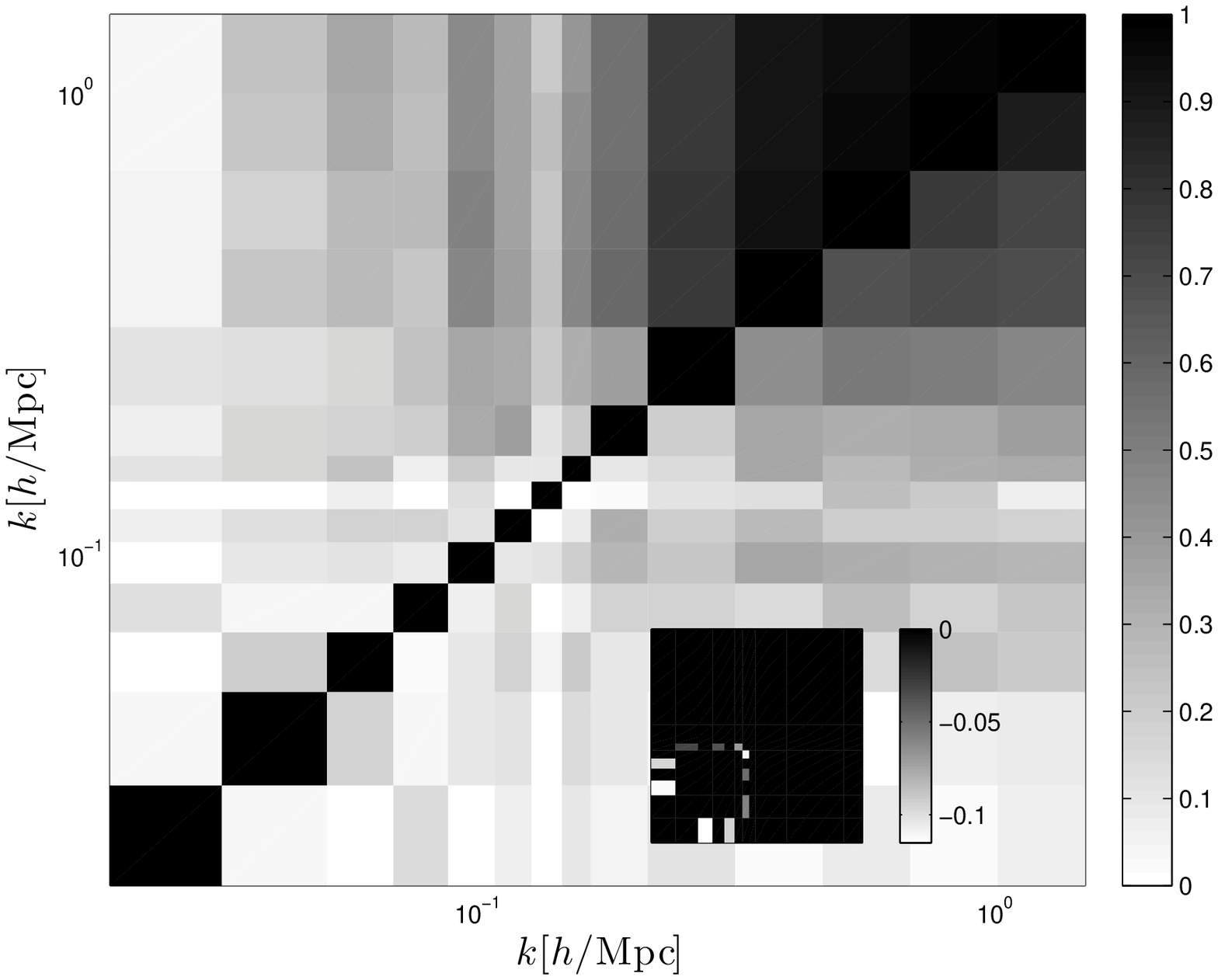}
    \includegraphics[width=3.2in]{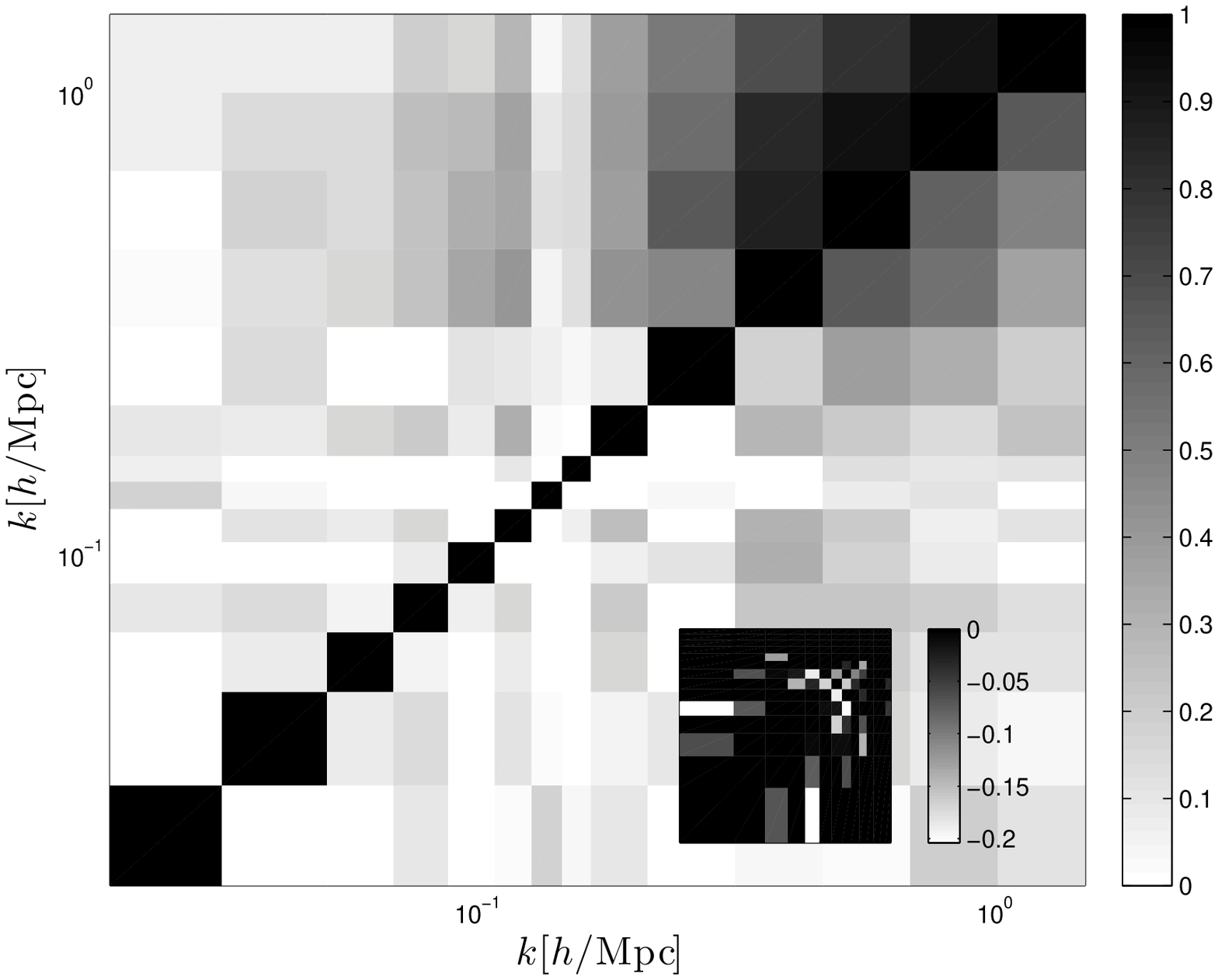}
    \includegraphics[width=3.2in]{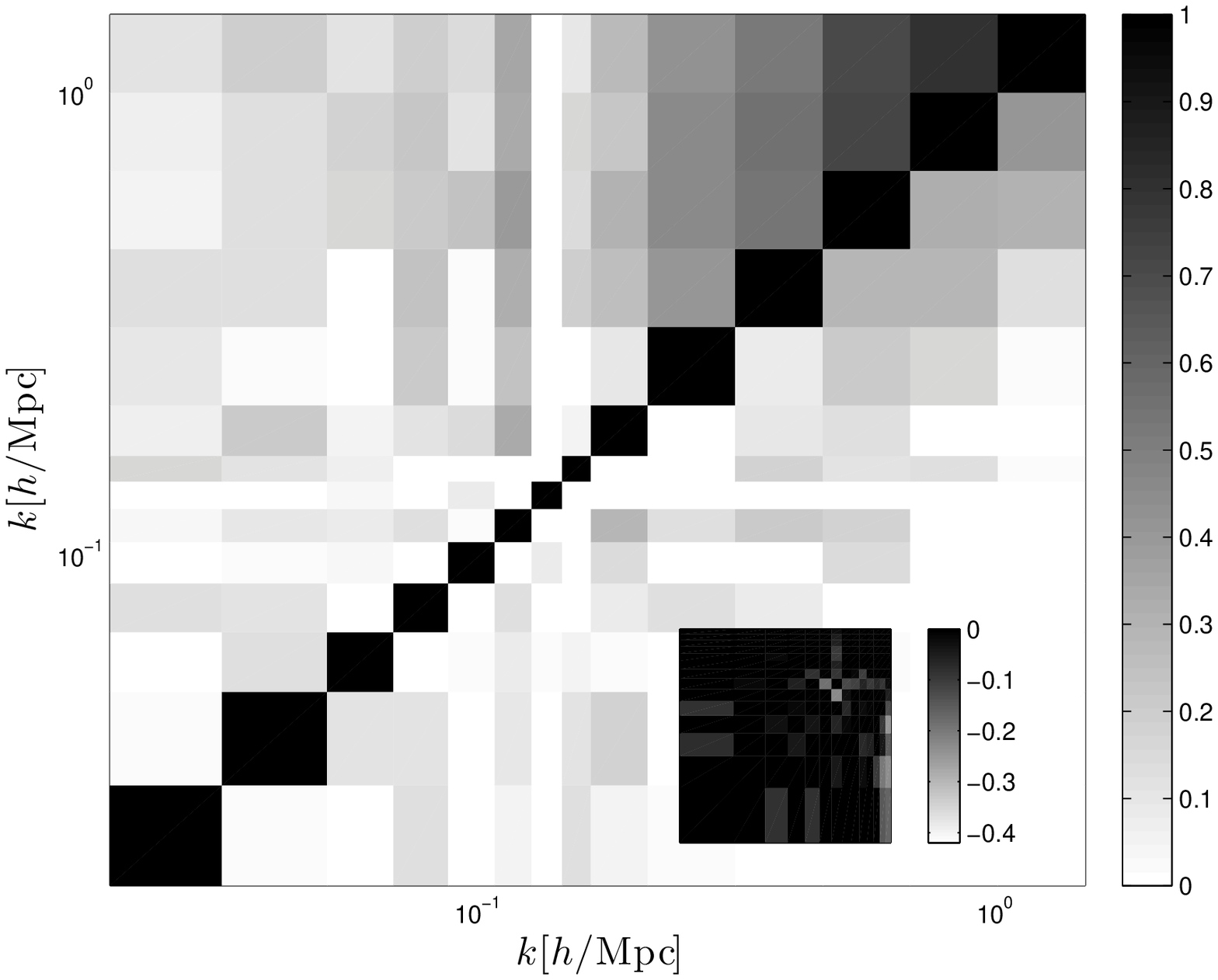}
    \includegraphics[width=3.2in]{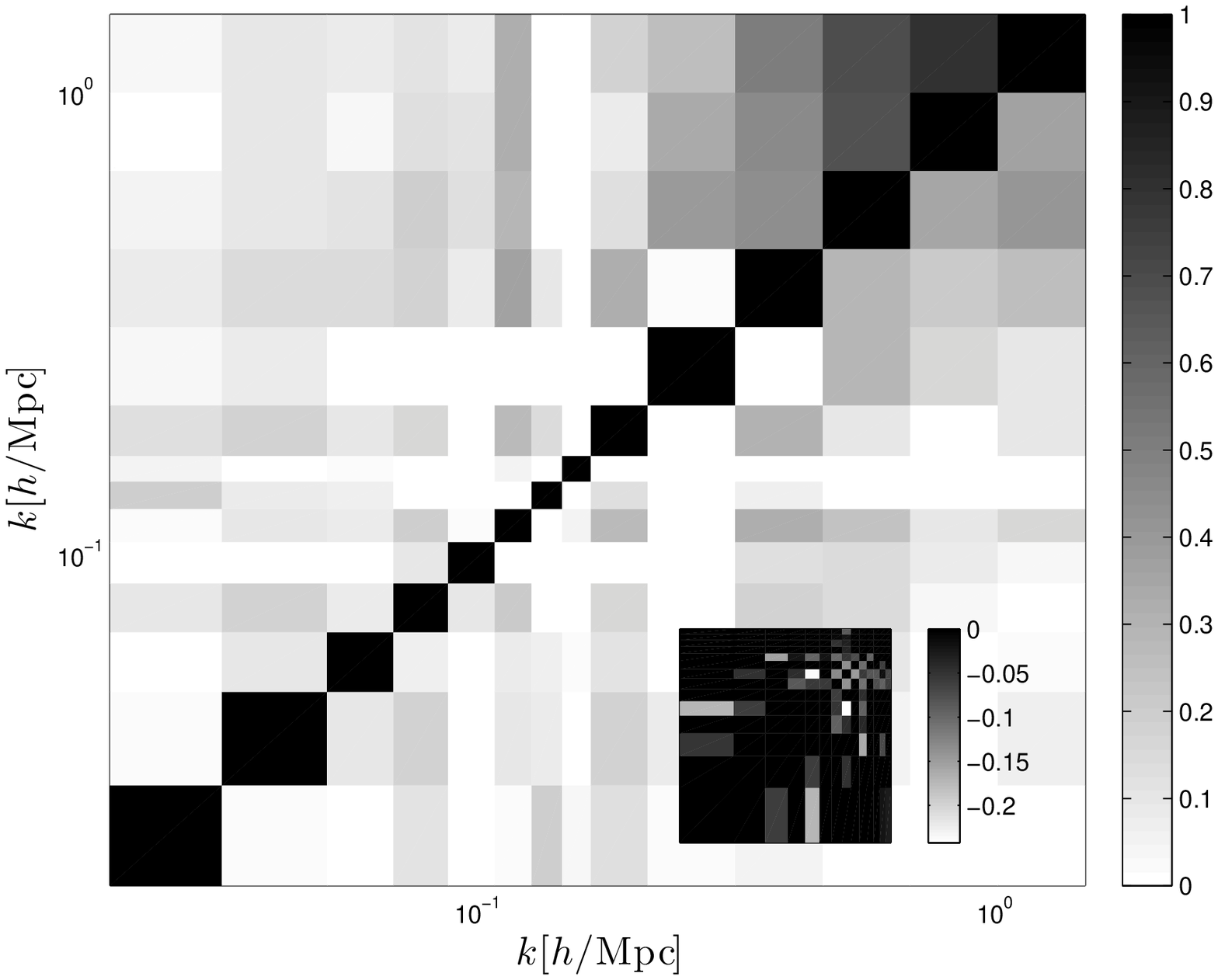}
     \caption{({\it top-left}:) Cross-correlation coefficient matrix associated with the particle power spectra. The top triangle represents measurements from the original matter fields, while the lower triangle elements are from fields after the NLWF has been applied. 
     The inset quantifies the amount of anti-correlation between the measurements. 
     ({\it top-right}:) The top triangle represents measurements from the reconstructed matter fields, while the lower triangle are from fields that are first reconstructed, then wavelet filtered -- still using particles as tracers. 
  ({\it bottom-left}:) 
 Cross-correlation coefficient matrix associated with the power spectrum measurements  from the simulated haloes. The top triangle represents measurements from the original halo fields, while the lower triangle are from fields that are  wavelet filtered. 
     ({\it bottom-right}:)  The top triangle represents measurements from the reconstructed halo fields, while the lower triangle are from fields that are first reconstructed, then wavelet filtered. }
\label{fig:rho_all}
\end{figure*}

The two Gaussianization techniques that are discussed in this paper both attempt to bring cosmological information, or degrees of freedom, back to the power spectrum. Consequently, the covariance matrices of the Gaussianized fields will be more diagonal. The top left panel of Fig. \ref{fig:rho_all} shows the cross-correlation coefficient matrix of the original particle fields in the upper triangle, and the wavelet filtered ones on the lower triangle. 
To ease the comparison between the figures, we show in the main figure of each panel the positive components only, and present in the insets the negative entries.
There is a mild anti-correlation (less than  10 per cent) in some matrix elements of the original fields, which comes from residual noise in the largest scale.
This is a mild effect that has very little impact, hence we do not attempt to correct for it. 
The top right panel shows, on the upper triangle, the results after a density reconstruction has been applied, then, on the lower triangle, the measurements after both technique have been executed.  The off-diagonal elements of the covariance matrix are reduced by 20-30 per cent by both Gaussianization techniques. We see that those two techniques combine well and reduce to a minimal value the correlation between the modes. 

The bottom panels of Fig. \ref{fig:rho_all} show the same measurements, when carried on halo fields. The wavelet filter produces a band of negative elements, correlating the $k>1.0 h\mbox{Mpc}^{-1}$ -- Poisson dominated -- with all scales. This anti-correlation does not carry any physics about the signal in it, hence these scales should be left out or carefully interpreted in future analyses. Most off-diagonal elements are about 30 per cent less correlated than in the unfiltered matrix, showing that wavelet filtering is also very efficient on halo fields. We also observe that the density reconstruction algorithm has very little impact on the correlation of the halo measurements. This is caused by the fact that haloes are non-overlapping by construction, hence the region of exclusion prevents an accurate construction of the gravitational potential\footnote{We could of course improve the performance of the reconstruction technique by using the gravitational potential measured from simulated particles, which we have at hand. Even in a data set, it is in principle possible to combine independent measurements of the potential, obtained say with weak lensing tomography. This is an interesting avenue that is, however, beyond the scope of this paper.}. This is a strong limit of the technique, as real galaxy data behave much more like halos than particles. However, higher multiplicity in the galaxy population of many haloes is likely to improve the construction of the gravitational potential, hence we can expect a modest gain. When we combined both technique, most of the diagonalization comes from wavelet filter. This can be seen visually by comparing the lower triangles of both bottom panels.

\subsection{Fisher information}
\label{ss.fisher2}

\begin{figure*}
    \includegraphics[width=5.2in]{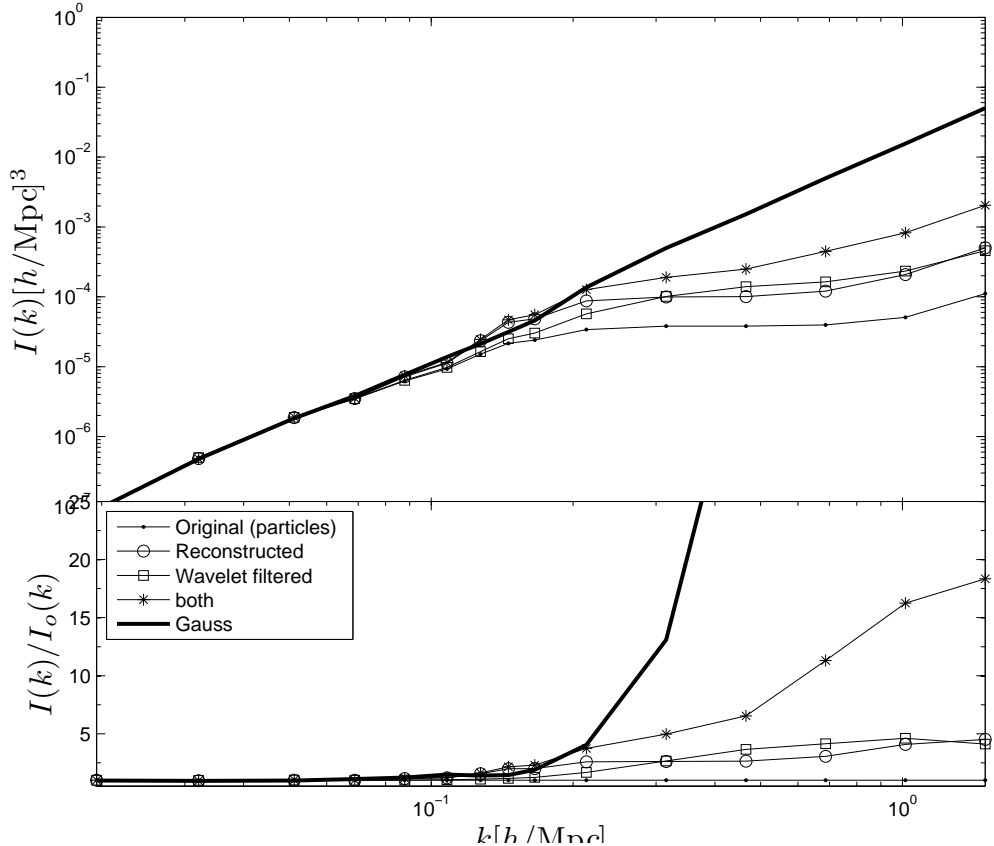}
     \caption{({\it top} :) Cumulative information contained in the dark matter power spectra of the original and Gaussianized fields from particles. As in Fig. \ref{f.power_pp}, the dots represent the original fields, the open circles show the results after our density reconstruction algorithm, the squares correspond to the wavelet filtered fields, and the stars represent a combination of both techniques. The analytical Gaussian (i.e. linear) Fisher information curve is shown with the thick solid line. These two Gaussianization techniques are shown to work in conjunction, such that on all scales, their combined effect recovers the largest amount of information. For $k>0.6h \mbox{Mpc}^{-1}$, the improvement on particles is more than an order of magnitude. ({\it bottom} :) Ratio of the lines presented in the top panel with the original fields.}
    \label{f.info_pp}
\end{figure*}

When we extract the Fisher information from the covariance matrices presented above, we expect the original particle fields to exhibit the global shape first measured in \cite{2005MNRAS.360L..82R}.
Namely, the information should follow the Gaussian predictions on large scales, then reach a trans-linear plateau where the gain is very mild as one increases the resolution of the survey, then hit a second rise on scales smaller than about $1.0 h\mbox{Mpc}^{-1}$. We first see from Fig. \ref{f.info_pp} that we are able to recover those results, plus those of \cite{2011arXiv1106.5548N}, which showed that the density reconstruction algorithm can raise the height of the trans-linear plateau by a factor of a few. We also recover the results from \cite{2011ApJ...728...35Z} and obtain a similar gain with the wavelet non-linear Wiener filtering technique.

As mentioned in the introduction, it was shown  by \cite{2012MNRAS.421..832Y} that different Gaussianization techniques do not always combine well. In the current case, however, we observe that on all scales, the Fisher information from the combined techniques are larger than the sum of the two separate contributions. For  $k > 0.6 h\mbox{Mpc}^{-1}$, notably, we are able to extract more than ten times the Fisher information of the original particles fields, whereas individual techniques offer a recovery of about a factor of four. This symbiosis effect grows larger as one goes to smaller scales.

\begin{figure*}
    \includegraphics[width=5.2in]{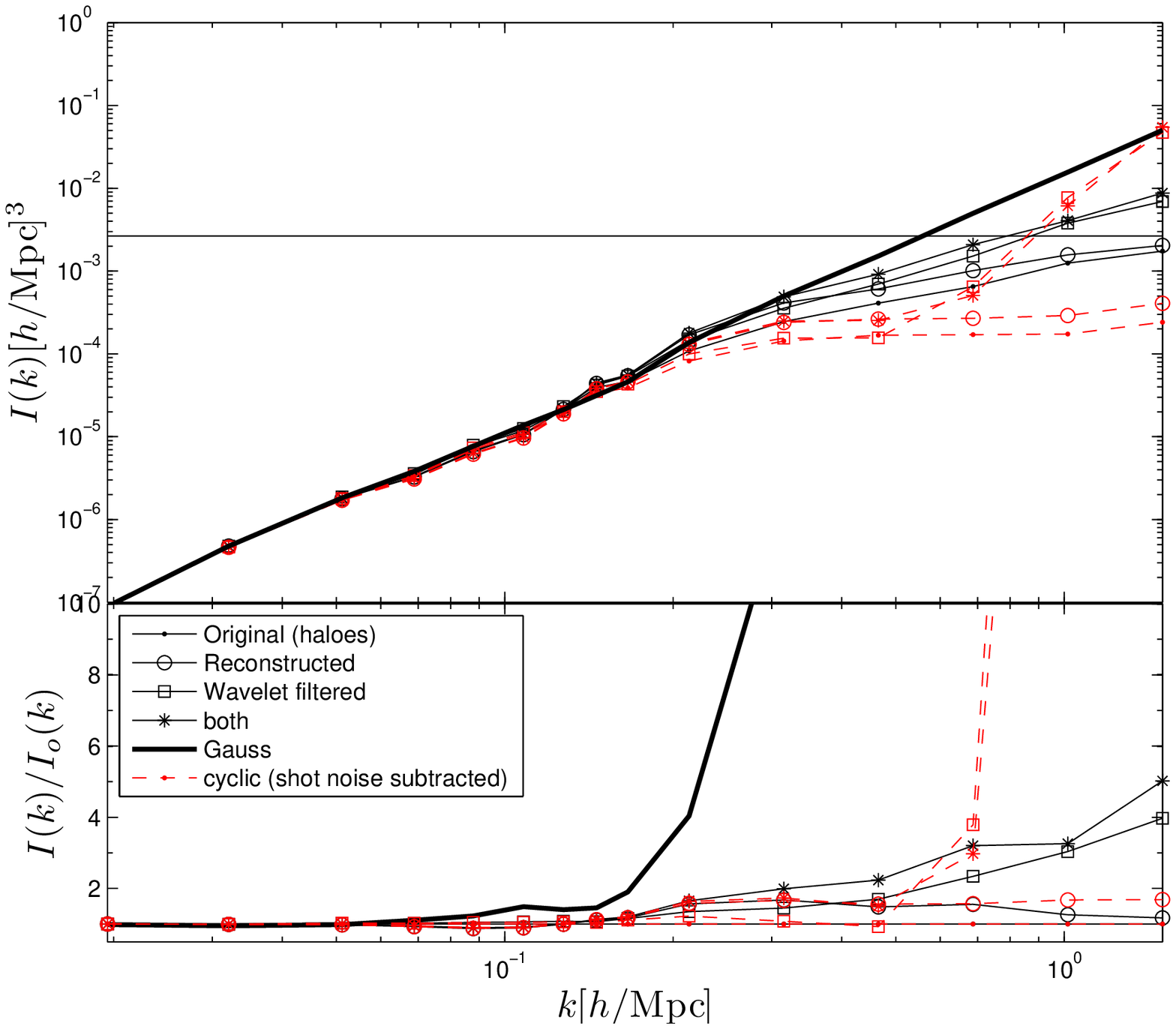}
     \caption{({\it top} :) Cumulative information contained in the dark matter power spectra of the original and Gaussianized fields from haloes. The analytical Gaussian (i.e. linear) Fisher information is shown with the thick solid line. Reconstruction offers only a mild improvement on the Fisher information when taken alone, whereas wavelet filter recovers three times more information by $k=0.7h \mbox{Mpc}^{-1}$. The flat line corresponds to the halo number density, and the dashed lines are for shot noise subtracted calculations (in red on the on-line version). We observe that the (shot noise included) halo information saturates at the number density, as expected from non-Gaussian Poisson density fields. Wavelet filtered halo densities have a lower shot noise, hence can exceed this Poisson limit. ({\it bottom} :) Ratio of the lines presented in the top panel with the original fields. We see that shot noise subtracted fields shows a milder information recovery. This due to the presence of non-Gaussian Poisson noise in the original fields, which was largely removed by the wavelet filter, thus boosting the performance.}
    \label{f.info_halo}
\end{figure*}

When considering the halo fields, we observe in Fig. \ref{f.info_halo} that the density reconstruction technique, taken alone, has little impact on the recovery of information, due to a poor modeling of the gravitational potential. In contrast, wavelet filtering recovers five times more information by the time we have reached $k=1.0h \mbox{Mpc}^{-1}$, before shot noise subtraction. The Poisson noise is a non-Gaussian effect, which also saturates the Fisher information. A hard limit one can think of is the following: the number of degrees of freedom can not exceed the number of objects in our fields of view. We therefore plot the (non-Gaussian) Poisson noise limit as a flat line corresponding to the halo number density, and observe that the original halo Fisher information approach but never exceed that limit. Wavelet fields, however, reduces the Poisson noise significantly, hence allows the information to reach higher values. We also see that shot noise subtracted Fisher information curves show a lower information recovery, which means that the number density needs to be high enough in order to maximize the recovery.
As mentioned in section \ref{ss.sims}, the halo density is about eight times larger than current spectroscopic surveys.
Next generation experiments and current photometric redshift surveys have a much larger number counts, 
hence the corresponding Poisson noise limit will be much higher.

\section{Discussion and Conclusion}\label{s.conclusion}

This paper explores the recovery of Fisher information with the combined use of two Gaussianization techniques, and show that wavelet non-linear Wiener filtering and density reconstruction can extract an order of magnitude more information than the original fields. We also reproduce the calculations on halo catalogues and find that 1) the density reconstruction has only a mild impact on its own, due to a poor modelling of the gravitational potential, 2) wavelet filter recovers about five times more information by $k=1.0h \mbox{Mpc}^{-1}$, and 3) the combined techniques recovers about three times more Fisher information than in the original fields by $k>0.7h \mbox{Mpc}^{-1}$, even after shot noise subtraction. Interestingly, we find that in both matter tracers, the recovery of the combination of the two Gaussianization techniques is larger than the sum of the individual contributions.

\section*{Acknowledgements}
This work was supported by the National Science Foundation of China (Grants No. 11173006), the Ministry of Science and Technology National Basic Science program (project 973) under grant No. 2012CB821804, and the Fundamental Research Funds for the Central Universities. Computations were performed on the TCS supercomputer at the SciNet HPC Consortium. SciNet is funded by: the Canada Foundation for Innovation under the auspices of Compute Canada; the Government of Ontario; Ontario Research Fund - Research Excellence; and the University of Toronto. UP and JHD would like to acknowledge NSERC for their financial support.

\bibliographystyle{mn2e}
\bibliography{haoran_ref}
\label{lastpage}

\end{document}